\documentclass[times,twocolumn,final]{elsarticle}

\usepackage{framed,multirow}
\usepackage{algorithm}
\usepackage{algpseudocode}
\usepackage{amssymb}
\usepackage{latexsym}
\usepackage{amsmath}
\usepackage{booktabs}
\usepackage{url}
\usepackage{xcolor}
\usepackage{ulem}
\usepackage{hyperref}

\definecolor{newcolor}{rgb}{.8,.349,.1}

\begin{document}

\begin{frontmatter}

\title{Narrowing the semantic gaps in U-Net with learnable skip connections: The case of medical image segmentation}%

\author[1,2]{Haonan Wang}
\ead{haonan1wang@gmail.com}
\author[1,2]{Peng Cao \corref{cor1}}
\cortext[cor1]{Corresponding author: 
  E-mail address: caopeng@mail.neu.edu.cn.}
\author[3]{Xiaoli Liu}
\author[1,2]{Jinzhu Yang}
\author[4]{Osmar Zaiane}

\address[1]{School of Computer Science and Engineering, Northeastern University, Shenyang, China}
\address[2]{Key Laboratory of Intelligent Computing in Medical Image of Ministry of Education, Northeastern University, Shenyang, China}
\address[3]{DAMO Academy, Alibaba Group, Hangzhou, China}
\address[4]{Amii, University of Alberta, Edmonton, Canada}


\begin{abstract}
Most state-of-the-art methods for medical image segmentation adopt the encoder-decoder architecture. 
However, this U-shaped framework still has limitations in capturing the non-local multi-scale information with a simple skip connection. To solve the problem, we firstly explore the potential weakness of skip connections in U-Net on multiple segmentation tasks, and find that i) not all skip connections are useful, each skip connection has different contribution; ii) the optimal combinations of skip connections are different, relying on the specific datasets. Based on our findings, we propose a new segmentation framework, named UDTransNet, to solve three semantic gaps in U-Net. Specifically, we propose a Dual Attention Transformer (DAT) module for capturing the channel- and spatial-wise relationships to better fuse the encoder features, and a Decoder-guided Recalibration Attention (DRA) module for effectively connecting the DAT tokens and the decoder features to eliminate the inconsistency. Hence, both modules establish a learnable connection to solve the semantic gaps between the encoder and the decoder, which leads to a high-performance segmentation model for medical images.
Comprehensive experimental results indicate that our UDTransNet produces higher evaluation scores and finer segmentation results with relatively fewer parameters over the state-of-the-art segmentation methods on different public datasets. Code: \underline{https://github.com/McGregorWwww/UDTransNet}.
\end{abstract}


\end{frontmatter}



\section{Introduction}
Medical image segmentation and the subsequent quantitative assessment of target objects~\citep{UNet,AND,wang2023dhc,cao2022collaborative} provide valuable information for the disease diagnosis and treatment planning.
Recent semantic segmentation approaches \citep{InfNet, NENet, MetricUNet, CHEN2022102311, FAT-Net}  typically rely on a U-Net like encoder-decoder architecture \citep{UNet} where the encoder produces high-level semantic features, the decoder upsamples these hidden features gradually to produce segmentation maps with per-pixel probability
and the skip connections incorporate the contextual information through the encoding-decoding process by transiting the feature maps from the encoder to the decoder.

\begin{figure}[t]
	\centering
	\includegraphics[width=\linewidth]{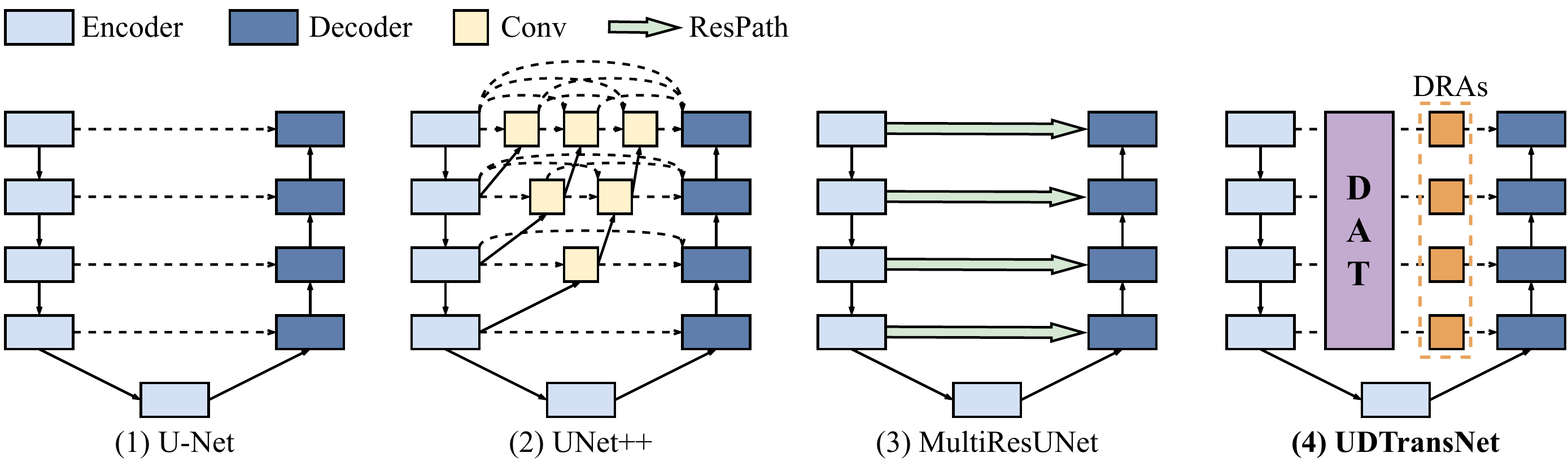} 
	\caption{Comparison of the different skip connection schemes. }
	\label{U_skip}
\end{figure}
The skip connection scheme in the U-shaped framework still has limitations in capturing the non-local multi-scale information, since it simply copies the features from the encoder to the decoder.
To investigate the effectiveness of the skip connections, we conducted an in-depth study and found that: 
1) not all skip connections are useful, each skip connection has different contribution; 
2) the optimal combinations of skip connections are different, relying on the specific datasets.
The underlying reasons {may be the} semantic gaps between the encoder and the decoder stages.
To overcome these limitations, several previous methods have been developed to alleviate the inconsistency by introducing additional convolution operations on the features transmitted from the encoder, such as UNet++~\citep{UNet++} and MultiResUNet~\citep{MultiResUNet}.
We illustrate the architectures to show their differences in Fig.~\ref{U_skip}.
In spite of achieving good segmentation results, these methods above still have limitations in sufficiently exploring the full-scale information.
Another issue is that although the gap is alleviated by some convolution process, it cannot guarantee the transited features are compatible with the decoder.

\begin{figure}[t]
	\centering
	\includegraphics[width=\linewidth]{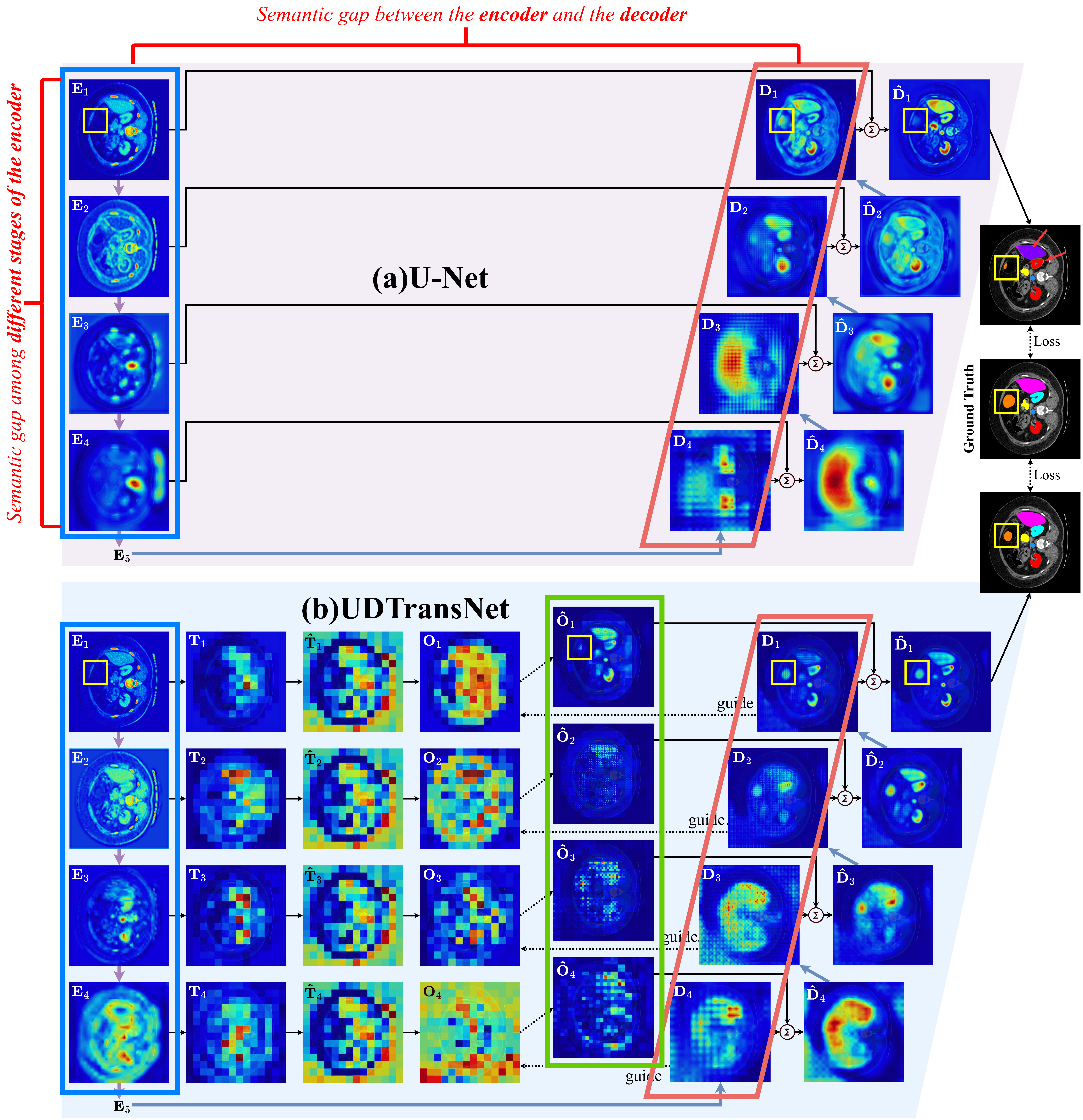} 
	\caption{Illustration of the semantic gaps in U-Net and how our proposed UDTransNet solves the gaps. Blue, red and green boxes indicate the feature maps of original encoders, decoders and the outputs of  DAT, respectively;
 {yellow boxes} indicate the most semantically inconsistent areas which lead to unexpected results of U-Net.
 }
	\label{multi_scale}
\end{figure}

In this work, we suggest two key aspects of potential semantic gaps crucial to be considered: \textit{the semantic gap among multi-scale features in different encoding stages} and \textit{the semantic gap between the encoder and the decoder}. The semantic gaps bring severe feature inconsistency, which can clearly be observed through the feature maps in Fig.\ref{multi_scale}(a).
Driven by the semantic gaps, two critical questions arise: how to balance the semantic gaps among the multi-scale multi-channel feature maps in the encoding stage, and how to effectively balance the semantic gaps between the encoder and the decoder?
Our work aims to find well-informed answers to these questions.
To this end, we develop an end-to-end segmentation network named UDTransNet, which is based on the U-shaped structure and consists of three attention-based modules. 
More specifically, the semantic gap among different encoding stages is mainly solved by  the proposed DAT module, which is composed of the  Channel-wise Fusion Attention (CFA) sub-module and the Spatial-wise Selection Attention (SSA) sub-module. Concretely, the CFA sub-module \textit{concatenates the four encoder features} of all levels and leverages the self-attention operation for modeling the sophisticated channel-wise dependencies to align the channels in multiple stages. 
Subsequently, the SSA sub-module with \textit{cross-attention mechanism} fuses the multi-stage encoding features and highlights the important regions across all scales by modeling spatial-wise long-range dependencies.
The features among the encoder-decoder stages are more consistent after the CFA and SSA sub-modules, as can be seen in the third and fourth rows in Fig.~\ref{multi_scale}(b).
The DAT module mitigates the semantic gap between the encoder and the decoder to some extent through global self-attention mechanisms. Nevertheless, it is not sufficient due to the inconsistency of the  self-attention operations in DAT and the convolution-based operation in the encoder-decoder architecture. Thus, we further propose the Decoder-guided Recalibration Attention (DRA) module to recalibrate the outputs of DAT and obtain more appropriate features for the decoder.
As shown in Fig.~\ref{multi_scale}(b), compared with the encoder features (in the blue box), the features after DAT and DRA are more consistent with the decoder features (in red box), and some important features are highlighted, \textit{e.g.}, the regions highlighted by the yellow boxes.

The main contributions of this paper are as follows. \\
(1) We made a thorough analysis to investigate the effectiveness of the skip connections in U-Net on multiple datasets.\\
(2) {Instead of incorporating Multi-head Self Attention (MSA) into the encoder-decoder architecture \citep{SwinUnet} or only encoder stage \citep{TransUNet}}, we propose UDTransNet and argue that employing the well-designed attention modules to serve as strong skip connection is a more appropriate manner {for boosting multi-scale feature fusion, bridging the potential semantic gaps and reducing the amount of trained parameters.}
Moreover, they are flexible and can be easily plugged into various segmentation models.\\
(3) Extensive experiments on {five} public datasets verify that our method not only achieves excellent performance, but also leads to fewer computational costs compared with the state-of-the-art methods.

This paper extends our previous conference paper \citep{UCTransNet} in several aspects. \\
\textbf{(1)} The \textbf{architecture}  of UDTransNet in this journal paper is significantly improved as follows:
\textit{(i)} the multi-scale features learned by the CCT (Channel-wise Cross Fusion Transformer) module in UCTransNet are prone to be similar, and  the spatially local information is difficult to be reflected by the channel features.
Hence, we replace it with a combination of CFA and SSA modules to model both channel-wise and spatial-wise dependencies across scales to learn meaningful representations;
\textit{(ii)} the CCA (Channel-wise Cross Attention) module in UCTransNet is a local convolution operation, which is inconsistent with the global dependencies learned by Transformer when fused with decoder features, thereby we replace it with a matrix-based decoder-guided recalibration attention (DRA) module to alleviate the discrepancy when fusing these two incompatible feature sets.
In summary, the proposed modules are pure multi-head attention components in the overall framework except {for} the convolution operations in the vanilla encoder-decoder stages.\\
\textbf{(2)} Finally, we conduct more comprehensive \textbf{experiments} to validate the effectiveness of our methods based on our previous work~\citep{UCTransNet}:
\textit{(i)} in addition to the three datasets used in the previous paper, the proposed UDTransNet is further evaluated on two challenging datasets with large {amounts} of images: ISIC skin lesion segmentation dataset with vague boundaries and ACDC multi-class MRI cardiac segmentation dataset with complex object structures;
\textit{(ii)} verifying the introduced semantic gaps by visualizing the feature maps;
\textit{(iii)} more comprehensive and convincing segmentation visualizations;
\textit{(iv)} more ablation study of different settings in the DAT module to systematically investigate different settings of the proposed DAT module and verify the different roles of the three proposed attention-based modules in UDTransNet;
\textit{(v)} systematically investigating different factors on the segmentation performance of UDTransNet such as the numbers of layers and heads.


\section{Related Works}

\subsection{Skip Connections in U-shaped Nets}

The skip connection is designed to help transmit the spatial information that gets lost during the encoder stages to help recover the {high-resolution} information. It is unnecessarily restrictive as it demands the feature fusion of the encoder and decoder at the same level without considering the semantic gap.
{To solve the issue, some U-Net methods with different skip connections for improved performance are proposed. These solutions can be classified into two main categories: using dense connection designs between the encoder and decoder~\citep{UNet++,MultiResUNet,qian2022unet} and attention mechanisms~\citep{AttentionUNet,taghanaki2019select,jia2022learning}.}
Specifically, for the first category, \cite{UNet++} suggested that the features from the same scale of the encoder and decoder are semantically inconsistent and thus developed a nested framework named UNet++ to bridge the gap by capturing multi-scale features.
{UNet\#~\citep{qian2022unet} aims to further improve the skip connections in UNet++ with more dense connections by aggregating feature maps of different scales in the decoder, so as to capture fine-grained details and coarse-grained semantics from the full scale.}
In addition, MultiResUNet \citep{MultiResUNet} is proposed to solve the semantic gap by introducing the ResPath module with the residual structure for enhancing the skip connections.

{
For the attention-based category, Attention U-Net~\citep{AttentionUNet} is designed with a cross-attention module with decoder features as gating signals to disambiguate irrelevant responses in skip connections. 
\cite{taghanaki2019select} proposed select-attend-transfer gate to select  the most discriminative feature channel. It focuses on  learning to identify the most discriminative feature maps in a skip connection through a convolutional filter and  an attention layer from the channel-wise perspective.
\cite{jia2022learning} also proposed a channel-wise attention mechanism, combined with cascaded pyramid convolutional block  to exploit the multi-scale spatial contextual information.}


{
Though these methods focusing on the skip connections have shown promising results, they still suffer from the following three major limitations: 1) they are CNN-based approaches, which suffer from a limitation in modeling the long-range semantic dependencies due to a confined receptive field size; 2) most work  adopts only the channel-wise attention, lacking the simultaneous exploitation of the channel and  spatial relationship of skip connections related to the semantic gaps; and 3) they fail to exploit the multi-scale fusion.
}

\begin{figure}[t]
	\centering
	\includegraphics[width=0.9\linewidth]{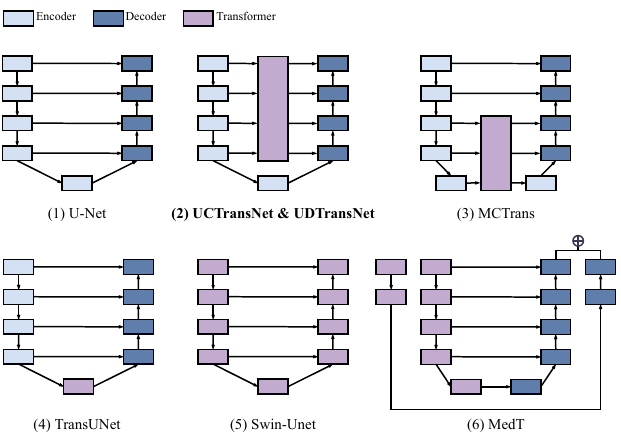} 
	\caption{The architecture of different transformer-based methods.}
	\label{topo_graph}
\end{figure}

\subsection{Transformers for Medical Image Segmentation}
{Modeling the long range dependencies is essential in medical image segmentation. Early methods, involving convolution-based models~\citep{cbam,AttentionUNet}  or simple attention-based models~\citep{nonlocal, nonlocalunet}, enlarge the receptive fields  but still fail to establish effective long-range dependencies.}
In recent years, Vision Transformer (ViT) \citep{ViT} has achieved state-of-the-art on the computer vision domain by redesigning the transformer originally proposed in natural language processing.
{Transformers successfully model the long-range dependencies with 1) multi-head scheme of Self-Attention (MSA), 2) stacked layer normalizations and MLPs, and 3) positional encodings, and obtain large performance improvements compared with the traditional attention schemes. Thus, several researchers have incorporated transformers into medical image segmentation tasks \citep{RethinkingSemanticSegmentation,MultiBranch, UTNet,TransFuse,UNETR}.}
TransUNet \citep{TransUNet} is the first work that integrates transformers on medical image segmentation tasks. It adds ViT at the end of the encoder stage to improve the feature learning capability.
{MedT \citep{MedT} is designed with a gated position-sensitive axial attention mechanism and consists of four gates to control the amount of information which is supplied to the key, query, and value of transformer by the positional embedding.}
Motivated by Swin Transformer \citep{Swin}, Swin-UNet \citep{SwinUnet} is proposed by replacing the convolution blocks with the Swin Transformer modules in U-Net.
Moreover, MCTrans \citep{MCTrans} is proposed to capture cross-scale contextual dependencies among different encoder stages via a self-attention module and learns the correspondence among different classes via a cross-attention module.
Fig.~\ref{topo_graph} shows the architecture comparison of different transformer-based segmentation methods.


Most of the recent transformer-based segmentation methods mainly focus on improving the encoder by capturing long-range dependencies. 
{These methods incorporate the transformer into U-Net in a simple way, i.e. directly applying ViT as the encoder after the convolution operations. }
However, through the analysis in Section~\ref{section:analysis}, we argue that the major limitation of the current {U-shaped segmentation models is the issue of the skip connections rather than the encoder. 
On the one hand, exploring the multi-scale context is crucial for improving the segmentation performance. 
On the other hand, narrowing the semantic gap between the encoder and the decoder with learnable skip connections should be taken into consideration for improving the
segmentation performance.}

\begin{figure}[t]
	\centering
	\includegraphics[width=\linewidth]{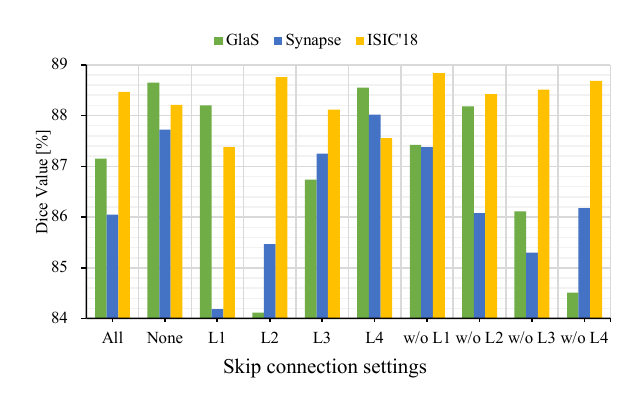} 
	\caption{Comparison of different skip connection settings in U-Net {with respect to the mean Dice values of 5-fold CV.} `None' represents the U-Net without any skip connection, `All' represents the vanilla U-Net, `L1' or `w/o' indicates only the skip connection of 1st level is kept or removed, respectively.
	}
	\label{skip}
\end{figure}

\section{The Analysis of Skip Connection } 
\label{section:analysis}

In this section, we thoroughly analyze the contributions of the skip connection to the performances on three datasets and obtain two findings highlighted as follows:


\begin{figure*}[t]
	\centering
	\includegraphics[width=\textwidth]{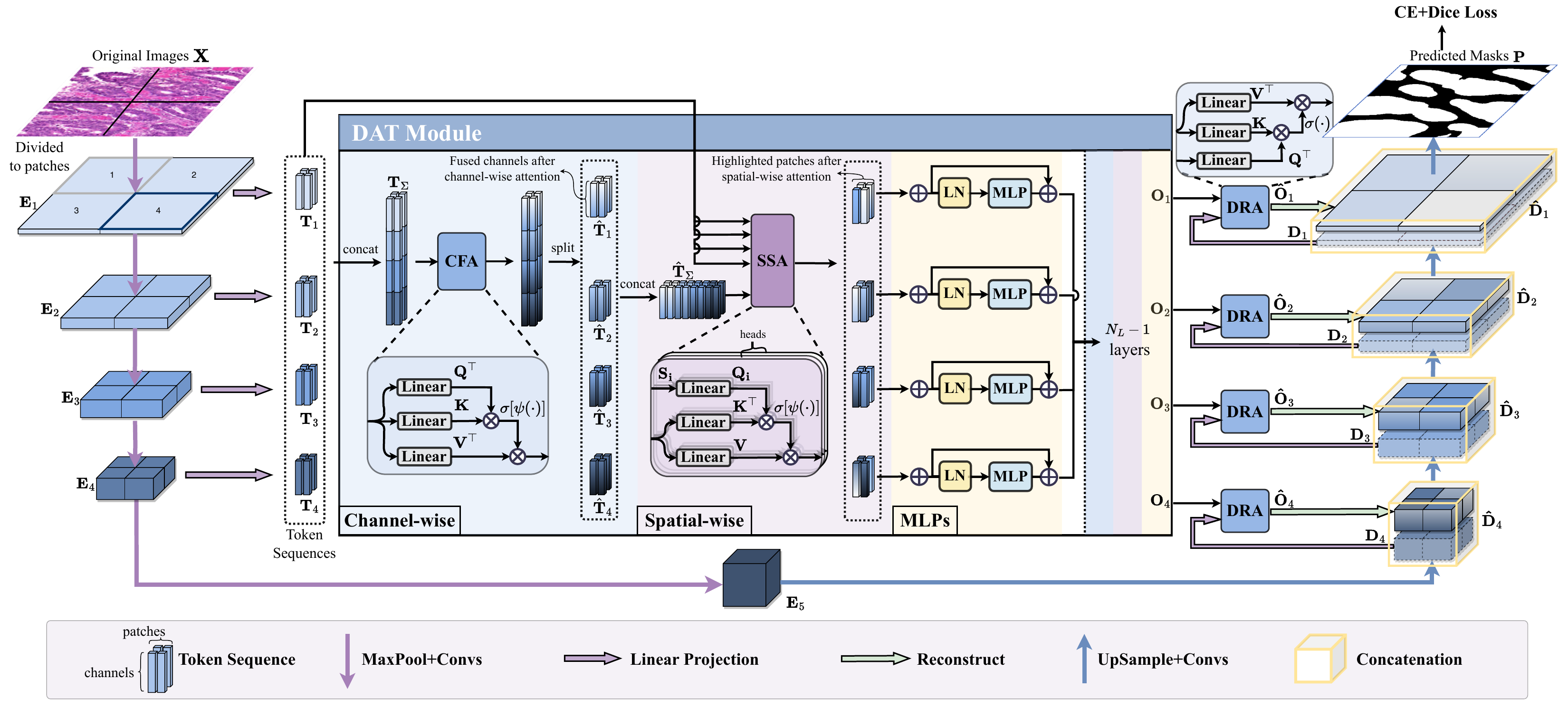} 
	\caption{Illustration of the proposed UDTransNet. We replace the original skip connections by two components: Dual Attention Transformer (DAT) and Decoder-guided Recalibration Attention (DRA) module. `LN' denotes Layer Normalization.}
	\label{framework}
\end{figure*}

\textit{Finding 1: Not all skip connections are useful, each skip connection has \textcolor{red}{a} different contribution.}
{From Fig.~\ref{skip}, we find `None', a U-Net without any skip connection is worse than `All', i.e. the vanilla U-Net, on the ISIC’18 (skin lesion segmentation) dataset, whereas it achieves very competitive performance against `All' on the GlaS (colon histology) dataset and the Synapse (abdominal segmentation) dataset.}
Moreover, the variation of dice value is large for different single skip connections, some skip connections even harm the segmentation performance.
The results imply that the original skip connections are not always beneficial for segmentation. The reason behind it {may} be that the simple connection is not appropriate for the segmentation tasks due to the incompatible features from the {stages of} the encoder and decoder.

\textit{Finding 2: The optimal combinations of skip connections are different, relying on the specific datasets.}
For example, `w/o $\mathrm{L}_1$' performs best on the ISIC dataset. However, it is surprising that `$\mathrm{L}_4$' with only one skip connection beats others on the Synapse dataset. 
These results indicate that the optimal combinations are different for different datasets, which further confirms the necessity of appropriately introducing learnable connections rather than simple copying.

\section{Methods}

\subsection{Framework}

Fig.~\ref{framework} illustrates an overview of our UDTransNet framework, which is mainly composed of two modules, including  Dual Attention Transformer (DAT) for capturing the channel- and spatial-wise encoder feature relationships, and decoder-guided recalibration attention (DRA) for fusing the semantically incompatible features of encoder and decoder stages.

\subsection{DAT: Dual Attention Transformer for Encoder Feature Transformation}

The deeper features encode high-level information whereas shallower features contain rich spatial information.
Driven by this, we develop a multi-scale feature fusion method for the segmentation of objects with varied scales by channel- and spatial-wise attention mechanisms. 
More specifically, we propose the DAT module to fuse the features from multiple stages in the encoder by leveraging the ability of the transformer in the modeling of the global context.
This approach takes the patches as tokens and aims to model both channel- and spatial-wise dependencies across scales to learn meaningful representations.
The DAT module consists of four stages: multi-scale feature embedding, Channel-wise Fusion Attention (CFA), Spatial-wise Selection Attention (SSA) and Multi-Layer Perceptrons (MLPs).  
Here, the CFA block calculates the self-attention between channels  to learn the channel relations of multiple encoder stages (scales), whereas the SSA block exploits the cross-attention among patches and focuses on learning the global spatial correlation across scales.

\begin{figure*}[t]
	\centering
	\includegraphics[width=\textwidth]{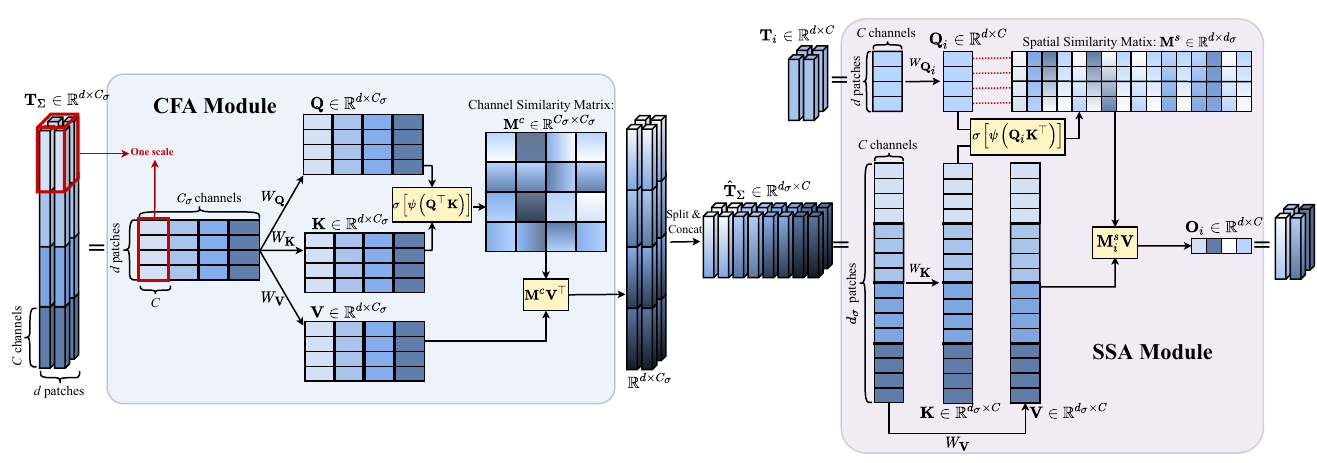} 
	\caption{Illustration of the learning procedure of the proposed CFA and SSA modules. Both modules incorporate  channel-wise and spatial-wise attention mechanisms to better learn the multi-scale representations from different aspects with global dependencies. Different scales are represented by different colors.}
	\label{op_process}
\end{figure*}

\subsubsection{Multi-scale Feature Embedding}
To enable the patches from multiple stages to focus on the  corresponding regions in the original image $\mathbf{X} \in \mathbb{R}^{H\times W}$, we firstly perform tokenization by reshaping the features from different scales $\mathbf{E}_i \in \mathbb{R}^{\frac{H}{2^{i-1}}\times \frac{W}{2^{i-1}} \times C_i},(i=1,2,3,4)$ into token sequences with patch sizes $P,\frac{P}{2},\frac{P}{4},\frac{P}{8}$.
The channel dimensions are all changed to $C=128$.
Then, we concatenate the token sequences of four scales $\mathbf{T}_i \in \mathbb{R}^{C \times d}$, $d=\frac{HW}{P^2}$ through channel-axis as the key and value $\mathbf{T}_{\Sigma} = \mathbf{concat}_{c}([\mathbf{T}_1,\mathbf{T}_2,\mathbf{T}_3,\mathbf{T}_4]), \mathbf{T}_{\Sigma} \in \mathbb{R}^{C_\sigma \times d}$, where $C_\sigma=4C$.


\subsubsection{Channel-wise Fusion Attention (CFA)}
Generally, different channels contain scale-specific semantic features, hence adaptively fusing these features is beneficial for complex medical image segmentation tasks.
To fuse the multi-scale information from different stages, we propose the CFA sub-module, a channel-wise attention operation along the channel-axis of the feature maps, which allows the encoder to learn the  relationship among channels and capture the global semantic dependencies.
The process is illustrated in Fig.~\ref{op_process}. 
Mathematically, we obtain a query, value and key after the linear layers as follows: 
\begin{equation}
\mathbf{Q}= \mathbf{T}_{\Sigma} W_{\mathbf{Q}},\mathbf{K}=\mathbf{T}_{\Sigma}W_{\mathbf{K}},\mathbf{V}=\mathbf{T}_{\Sigma}W_{\mathbf{V}}
\label{eq1}
\end{equation}
where $W_{\mathbf{Q}}, W_{\mathbf{K}}, W_{\mathbf{V}} \in \mathbb{R}^{C_\sigma \times d}$ are transformation  weights, $d$ is the sequence length (patch numbers) and $C_\sigma$ is the channel dimension of full scales ($C_\sigma = 4C$ in our method).
With $\mathbf{Q}, \mathbf{K}, \mathbf{V} \in \mathbb{R}^{C_\sigma \times d}$, the channel similarity matrix $\mathbf{M}^c \in \mathbb{R}^{C_\sigma \times C_\sigma}$ is produced and the value $\mathbf{V}$ is weighted by $\mathbf{M}^c$:
\begin{equation}
\mathbf{M}^c \mathbf{V^\top}  = 
\sigma\left[\psi\left( {\mathbf{Q}^\top\mathbf{K}}\right)\right] \mathbf{V}^\top 
\label{eq2}
\end{equation}
where $\psi(\cdot)$ and $\sigma(\cdot)$  denote the instance normalization~\citep{IN} and the softmax function.
Different from the original self-attention, CFA operates self-attention along the channel-axis rather than the patch-axis, then we utilize the instance normalization to normalize the similarity matrix for smoothly propagating the gradient.


\subsubsection{Spatial-wise Selection Attention (SSA)}
The output of CFA is split to $\hat{\mathbf{T}}_1, \hat{\mathbf{T}}_2, \hat{\mathbf{T}}_3, \hat{\mathbf{T}}_4$ and re-concatenated along the patch-axis $\hat{\mathbf{T}}_{\Sigma}=\mathbf{concat}_s([\hat{\mathbf{T}}_1,\hat{\mathbf{T}}_2,\hat{\mathbf{T}}_3,\hat{\mathbf{T}}_4])$, $\hat{\mathbf{T}}_{\Sigma}\in\mathbb{R}^{C \times d_\sigma}$ is used as the key and value of SSA, $d_\sigma = 4d$ in our method.
After CFA, the token sequences from all scales contain the global context information and consistent semantic information (see Fig.~\ref{multi_scale}). 
We further propose the SSA sub-module to learn spatial correlations among patches across multiple scales.  Given the tokens from all encoding stages, we consider the tokens from certain scales as queries to highlight the important spatial regions from the spatial-wise perspective. It allows our model to capture the spatial correlation between each scale and the full-scale information by exploiting the spatial correlation from multi scales.


More specifically, SSA has five inputs, including four token sequences $\hat{\mathbf{T}}_i$ as queries and a concatenated token sequence $\hat{\mathbf{T}}_{\Sigma}$ as key and value:
\begin{equation}
\mathbf{Q}_i= \hat{\mathbf{T}}_i W_{\mathbf{Q}_{i}},\mathbf{K}=\hat{\mathbf{T}}_{\Sigma}W_{\mathbf{K}},\mathbf{V}=\hat{\mathbf{T}}_{\Sigma}W_{\mathbf{V}}
\label{eq3}
\end{equation}
where $W_{\mathbf{Q}_{i}}\in\mathbb{R}^{C \times d}$, and $W_{\mathbf{K}}, W_{\mathbf{V}} \in \mathbb{R}^{C \times d_\sigma}$ are weights of different inputs.
The three feature representations are split into $N_H$ different sub-spaces, \textit{e.g.}, $\mathbf{Q}_i=[\mathbf{Q}_i^1,\mathbf{Q}_i^2,\dots,\mathbf{Q}_i^{N_H}]$. 
With $\mathbf{Q}_i^h\in\mathbb{R}^{\frac {C}{N_H} \times d}$ and $\mathbf{K}^h, \mathbf{V}^h\in\mathbb{R}^{\frac {C}{N_H} \times d_\sigma}$, the spatial-wise similarity matrices $\mathbf{M}_i^s$ are calculated and each token in $\mathbf{V}^h$ is weighted by $\mathbf{M}_i^s$ through a cross-attention (CA) mechanism:
\begin{equation}
\mathrm{\Theta}_i^h  = \mathbf{M}_i^s \mathbf{V}^h  = 
\sigma\left[\psi\left(\mathbf{Q}_i^h\mathbf{K}^{h\top} \right)\right] \mathbf{V}^h \ \  \in \mathbb{R}^{\frac {C}{N_H} \times d}    
\label{eq4}
\end{equation}
The outputs of subspaces are then concatenated to produce the final output $\mathbf{O}^{SSA}_i = \mathbf{concat}([\mathrm{\Theta}_{i}^1, \mathrm{\Theta}_{i}^2,\dots,\mathrm{\Theta}_{i}^{N_H}]), \in \mathbb{R}^{C \times d}$.
The process of SSA in $i$-th scale is shown in Fig.~\ref{op_process}.

\subsubsection{Multi-Layer Perceptrons (MLPs)}
{Following SSA, MLPs with residual structure are developed to encode channel dependencies for refining the features from  SSA. The output of each MLP is obtained as follow:}
\begin{equation}
\mathbf{O}_i = \mathbf{O}^{SSA}_i+ \mathrm{MLP}( \mathbf{Q}_i+\mathbf{O}^{SSA}_i)
\label{eq5}
\end{equation}
The layer normalization is omitted in the equation for simplicity. The operations in Eq.(\ref{eq1})-(\ref{eq6}) are repeated $N_L$ times to build a $N_L$-layer transformer. In our implementation, $N_H$ and $N_L$ are both set to $4$.
The outputs of the $N_L$-th layer, $\mathbf{O}_1$, $\mathbf{O}_2$, $\mathbf{O}_3$ and $\mathbf{O}_4$, are then passed into the following DRA module.

\subsection{DRA: Decoder-guided Recalibration Attention Module for Transformer Feature Recalibration in Decoder}
The decoder part aims to generate the segmentation maps relying on the learned features from the encoder stages. However, the importance of the decoder is always ignored in previous studies. How to fuse the feature transmitted by skip connections from the encoders is crucial for enhancing the decoder features.
As can be seen in Fig.~\ref{multi_scale}, there exists a semantic gap between the output of DAT $\mathbf{O}_i$ and the upsampled feature $\mathbf{D}_i$, which inevitably hinders the direct concatenation the features from DAT and the U-Net decoder.
Thus, in order to alleviate the discrepancy when fusing these two incompatible feature sets, we propose the DRA module to leverage the decoder features for guiding the channel-wise information filtration of the enhanced features obtained by the DAT and eliminating the inconsistency.
As such, the DRA module can be considered as {a} feature calibrator to automatically fuse the incompatible features.
{It can be easily applied to} any scenario where semantic gaps exist.

Mathematically, we take the $i$-th level DAT output $\mathbf{O}_i\in\mathbb{R}^{C \times d}$ and the $i$-th level decoder feature map $\mathbf{D}_i\in\mathbb{R}^{C \times H\times W}$ as the input.
Then, $\mathbf{D}_i$ is further converted to token sequence $\mathbf{T}_{\mathbf{D}_i}$ as query and $\mathbf{O}_i$ as key and value:
\begin{equation}
\mathbf{Q}_i= \mathbf{T}_{\mathbf{D}_i} W_{\mathbf{Q}_{i}},\mathbf{K}=\mathbf{O}_i W_{\mathbf{K}},\mathbf{V}=\mathbf{O}_i W_{\mathbf{V}}
\label{eq6}
\end{equation}
where $W_{\mathbf{Q}_{i}}\in\mathbb{R}^{C \times d}$, and $W_{\mathbf{K}}, W_{\mathbf{V}} \in \mathbb{R}^{C \times d_\sigma}$ are weights of different inputs.

Consequently, $\mathbf{K}, \mathbf{V}\in\mathbb{R}^{C \times d}$ are recalibrated with the guidance of $\mathbf{Q}_{i}\in\mathbb{R}^{C \times d}$ through cross-attention:
\begin{equation}
\tilde{\mathbf{M}}_i \mathbf{V}  = 
\sigma\left[\psi\left( {\mathbf{Q}_i^\top \mathbf{K}} \right)\right] \mathbf{V}^\top 
\label{eq7}
\end{equation}
where $\tilde{\mathbf{M}}_i$ are the channel-wise similarity matrices.

Finally, the outputs of DRA, $\hat{\mathbf{O}}_1$, $\hat{\mathbf{O}}_2,\dots,\hat{\mathbf{O}}_i$, are reconstructed through an up-sampling operation followed by a convolution layer and concatenated with the decoder features $\mathbf{D}_1$, $\mathbf{D}_2,\dots,\mathbf{D}_i$, respectively.
This design of DRA can preserve texture details in the decoding process by appropriately fusing the encoder features.
The top-level features $\mathbf{D}_1$ is then converted to the $K$-class predicted masks $\mathbf{P} \in \mathbb{R}^{H\times W \times K}$ using an upsampling layer and two convolution layers.

In summary, our UDTransNet uses matrix-multiplication-based attentions in three different ways:\\
\textbf{CFA}: the queries, the keys and values come from the concatenated features from  four scales  in encoder.\\
\textbf{SSA}: the queries come from  each scale feature in encoder, and the keys and values come from the CFA outputs across scales.\\
\textbf{DRA}: the queries come from the decoder feature while both the keys and values come from the multi-scale fused features obtained by SSA sub-module.

\section{Experiments}
\subsection{Datasets}
We choose five public datasets to evaluate our proposed model, including two small-scale dataset (GlaS and MoNuSeg), and three large-scale datasets (Synapse, ISIC-2018 and ACDC).

\textbf{GlaS Dataset for Gland Segmentation.}
Gland segmentation dataset (GlaS) \citep{GlaS} was used in Gland Segmentation in Colon Histology Images Challenge Contest held at MICCAI'2015.
The dataset consists \textcolor{red}{of} 85 training samples and 80 testing samples which are derived from 16 H\&E stained histological sections of colorectal adenocarcinoma. 

\textbf{MoNuSeg Dataset for Nuclear Segmentation.}
MoNuSeg dataset \citep{MoNuSeg} was used in Multi-organ Nuclei Segmentation Challenge Contest held at MICCAI'2018,
which has 30 images with 21,623
hand-annotated nuclear boundaries for training and 14 images for testing.
Each image was sampled from a separate whole slide image of H\&E stained tissue of several organs from The Cancer Genomic Atlas (TCGA).

\begin{table}[t]
\footnotesize
\caption{Details of the parameters and experiment settings.}
\resizebox*{\linewidth}{!}{
\begin{tabular}{l|ccccc}
\toprule
\multirow{2}{*}{Settings} & \multicolumn{5}{c}{Datasets}              \\ 
                          & MoNuSeg & GlaS & ACDC & Synapse & ISIC-18 \\ \midrule
image size              &   224      & 224  & 224  & 224    & 224   \\
learning rate             &  1e-3       & 1e-3 &  1e-4  & 1e-3    & 1e-4    \\
batch size                 &   4      & 4    &  24    & 24      & 24      \\ \midrule
patch size $P$                & 8        & 16   &  16    & 16      & 16      \\
channel dim $C$              & 128     & 128  & 128  & 128     & 128     \\
heads $N_H$                    & 4       & 4    & 4    & 4       & 4       \\
layers $N_L$                   & 4       & 4    & 4    & 4       & 4      \\ \bottomrule
\end{tabular}}
\label{settings}
\end{table}

\textbf{Synapse Dataset for Multi-class Abdominal Organs Segmentation.}
Synapse multi-organ segmentation dataset \citep{Synapse} was used in MICCAI'2015 Multi-Atlas Abdomen Labeling Challenge which consists of 12 cases (2211 axial CT images) for training and 12 cases (1568 images) for testing in 8 abdominal organs, e.g., aorta, gallbladder, spleen, etc.
Note that, rather than using the original splits in \citep{TransUNet}, we use five-fold cross validation in the training stage and calculate the average scores of five-fold models in the testing stage to make the results more stable and convincing, thus the results in Table~\ref{SOTA_Synapse} are different from the results in Swin-UNet~\citep{SwinUnet} and TransUNet~\citep{TransUNet}.

\begin{table*}[t]
\centering
\footnotesize
\caption{Comparison with the state-of-the-art methods on Synapse dataset. The average Dice and average HD95 (95\% Hausdorff distance) of the 5 folds are shown bellow. `SC' denotes the shallow CNNs and `Pre' denotes whether the model uses the pre-trained weights. For simplicity, we use abbreviations of the eight organs. AO: Aorta, GA: gallbladder, LK: left kidney, RK: right kidney, LI: liver, PA: pancreas, SP: spleen, ST: stomach. The best results are boldfaced.}
\resizebox*{\linewidth}{!}{
\begin{tabular}{@{ }l@{\ \ }ccc|c@{\ \ \ \ }c|c@{\ \ \ \ }c@{\ \ \ \ }c@{\ \ \ \ }c@{\ \ \ \ }c@{\ \ \ \ }c@{\ \ \ \ }c@{\ \ \ \ }c@{ }}
\hline
Method   &Param.  &GFLOPs &Inf. time  & Avg Dice $\uparrow$           & Avg HD95 $\downarrow$ &AO &GA &LK &RK &LI &PA &SP &ST\\ 
\hline
U-Net \citet{UNet}    &14.8 &50.3  &0.357     &84.60$\pm$2.44 &47.93$\pm$8.05 &91.2 &73.5 &87.6 &89.6 &87.9 &82.9 &83.6 &80.6  \\
R34-U-Net \citep{UNet}  & 24.6 & 55.6   & 0.361   &86.05$\pm$0.64	&36.32$\pm$2.49  &91.1 &72.4 &90.8 &86.6 &86.6 &86.4 &89.4 &84.8        \\
{U-Net w/ CBAM \citep{cbam}}   &34.9 &101.9 &0.368   &85.36$\pm$1.55 &34.91$\pm$3.65  &89.0 &76.4 &87.9 &86.2 &84.2 &85.3 &88.8 &85.6\\
UNet++ \citep{UNet++}    &74.5 &94.6  &0.376    &86.11$\pm$1.08	&35.88$\pm$2.15 &88.4 &76.6 &88.1 &88.1 &89.6 &85.6 &90.4 &87.1\\
Attention U-Net \citep{AttentionUNet}   &34.9 &101.9 &0.368   &82.21$\pm$5.84 &54.82$\pm$27.6  &80.8 &80.1 &84.4 &85.8 &81.2 &81.5 &80.7 &83.1\\
MultiResUNet \citep{MultiResUNet}  &57.2 &78.4  &0.425    &86.44$\pm$1.02  & 44.16$\pm$4.46 &\textbf{91.7} &74.5 &89.6 &87.2 &88.4 &86.0 &89.5 &84.4\\ \hline
MedT \citep{MedT}          &98.3 &131.5 &3.719    &85.02$\pm$3.31	&52.69$\pm$15.8  &85.7 &80.9 &88.2 &87.4 &85.6 &\textbf{87.3} &88.1 &85.5\\
TransUNet \citep{TransUNet} &105.0 &56.7  &0.453  &87.21$\pm$0.88 &34.04$\pm$1.85 &90.2 &75.4 &90.7 &89.5 &90.5 &85.5 &91.8 &83.7\\
Swin-UNet \citep{SwinUnet}.  &82.3 &67.3  &0.480    &86.73$\pm$0.67	&31.62$\pm$1.56 &88.2 &77.1 &89.0 &86.9 &89.8 &85.1 &90.3 &87.0\\
MCTrans \citep{MCTrans}  & 25.3  & 65.7 &0.452    &87.65$\pm$0.81 &31.24$\pm$2.85 &91.1 &76.4 &92.1 &91.0 &89.0 &86.0 &91.5 &83.8\\
UCTransNet \citep{UCTransNet}     & 65.6 &63.2 &0.477    &87.91$\pm$0.71 &31.45$\pm$2.45 &{91.5} &73.5 &{91.3} &89.8 &89.3 &\textbf{87.3} &93.0 &87.1\\ \hline
\textbf{UDTransNet}   & 33.8 &63.2 &0.396    & \textbf{89.28$\pm$0.58}	    & \textbf{24.63$\pm$1.69} &89.4 &\textbf{82.6} &\textbf{92.8} &\textbf{92.1} &\textbf{91.2} &85.8 &\textbf{93.3} &\textbf{87.1}\\
\hline
\end{tabular}}
\label{SOTA_Synapse}
\end{table*}

\begin{table*}[t]
\centering
\footnotesize
\caption{Quantitative results on the GlaS, MoNuSeg and ISIC datasets. All results are in `mean$\pm$std' format. The best results are boldfaced.
}
\begin{tabular}{@{ }l|cc|cc|cc@{ }}
\hline
\multirow{2}{*}{Method}    & \multicolumn{2}{c|}{GlaS} & \multicolumn{2}{c|}{MoNuSeg} & \multicolumn{2}{c}{{COVID-19}} \\ & Avg Dice $\uparrow$  & Avg HD95 $\downarrow$ & Avg Dice $\uparrow$  & Avg HD95 $\downarrow$ & Avg Dice $\uparrow$  & Avg HD95 $\downarrow$\\ 
\hline
U-Net \citep{UNet}        & 85.45$\pm$1.25     & 15.30$\pm$1.67 &76.45$\pm$2.62  &5.58$\pm$2.04 & 60.29$\pm$10.55     & 50.43$\pm$19.30\\
R34-U-Net \citep{UNet}   & 87.15$\pm$1.08     & 10.27$\pm$2.42  &77.70$\pm$2.31  &4.71$\pm$1.64 &65.33$\pm$8.19     & 38.19$\pm$9.76\\
UNet++ \citep{UNet++}        & 87.56$\pm$1.17     & 12.72$\pm$1.76 &77.01$\pm$2.10  &4.18$\pm$1.29 & 78.58$\pm$0.75     & 13.62$\pm$1.47\\
Attention U-Net \citep{AttentionUNet}      & 88.80$\pm$1.07	    & 9.10$\pm$2.48 &76.67$\pm$1.06	 &5.38$\pm$1.76    &   77.23$\pm$2.45	    & 16.21$\pm$1.38\\
MultiResUNet \citep{MultiResUNet}             & 88.21$\pm$1.05	    & 8.77$\pm$1.89 &78.22$\pm$2.47	 &3.34$\pm$0.97 &  77.64$\pm$1.75	    & 13.28$\pm$2.10\\ 
{Non-local U-Net \citep{nonlocalunet}}        & 88.83$\pm$0.63     & 9.27$\pm$1.06 &77.75$\pm$2.30  &3.84$\pm$1.43 & 77.31$\pm$1.32     & 12.65$\pm$1.52\\
{MSD-Net \citep{jia2022learning}   }     & 86.76$\pm$0.70     & 13.87$\pm$2.17 &76.42$\pm$1.21  &3.50$\pm$1.36 & 77.11$\pm$2.01     & 13.73$\pm$1.96\\
{UNet\# \citep{qian2022unet} }       & 88.03$\pm$1.12     & 11.67$\pm$1.52 &76.82$\pm$3.11  &4.66$\pm$1.81 & 78.55$\pm$0.79     & 11.60$\pm$1.83\\
{Inf-Net \citep{InfNet}}        & -     & - &-  &- & 78.93$\pm$0.37     & 10.62$\pm$1.19\\
{HistoSeg \citep{histoseg}}        & 87.25$\pm$1.19     & 12.43$\pm$1.58 &77.30$\pm$1.35  &3.13$\pm$1.95 & -     & -\\
\hline
MedT \citep{MedT}                & 85.92$\pm$2.93	    & 16.02$\pm$2.71 &77.46$\pm$2.38	 &6.74$\pm$2.65 & 78.87$\pm$1.78	    & 15.28$\pm$1.30\\
TransUNet \citep{TransUNet}    & 88.40$\pm$0.74	    & 9.00$\pm$1.17 &{78.53$\pm$1.06}	 &3.20$\pm$1.25 & 78.46$\pm$11.93	    & 12.23$\pm$1.03\\
Swin-UNet \citep{SwinUnet}      & 89.58$\pm$0.57	    & 8.27$\pm$1.64 &77.69$\pm$0.94	 &2.77$\pm$0.84 & 77.93$\pm$2.84	    & 14.73$\pm$1.88\\
MCTrans \citep{MCTrans}   & 88.44$\pm$0.74	    & 9.96$\pm$1.62   &77.87$\pm$3.04 &3.74$\pm$0.97  &79.70$\pm$0.38 &10.95$\pm$1.37\\
UCTransNet \citep{UCTransNet}       & 90.18$\pm$0.71    & {7.52$\pm$1.38} &79.08$\pm$0.67	 &{2.72$\pm$0.57} & 79.29$\pm$0.70	    & 10.82$\pm$1.37\\ \hline
\textbf{UDTransNet}   & \textbf{91.03$\pm$0.56}	    & \textbf{6.71$\pm$0.99} &\textbf{79.47$\pm$0.80} & \textbf{2.35$\pm$0.25} & \textbf{79.80$\pm$0.73}	    & \textbf{10.23$\pm$0.62}\\

\hline
\end{tabular}
\label{SOTA_GlaS}
\end{table*}

\begin{table}[t]
\centering
\footnotesize
\caption{Quantitative results on the validation set of ISIC-18 dataset.}
\begin{tabular}{@{ }lcc@{ }}
\hline
Method            & Avg Dice $\uparrow$           & Avg HD95 $\downarrow$ \\ 
\hline
U-Net                    & 88.15$\pm$0.27     & 14.57$\pm$0.81\\
R34-U-Net           & 88.47$\pm$0.62     & 13.67$\pm$0.76\\
UNet++          & 88.97$\pm$0.38     & 12.31$\pm$0.91\\
Attention U-Net      & 89.05$\pm$0.37	    & 13.34$\pm$0.58\\
MultiResUNet       & 89.20$\pm$0.23	    & 12.70$\pm$0.57\\  \hline
MedT                   & 87.67$\pm$0.37	    & 15.28$\pm$1.30\\
TransUNet       & 89.32$\pm$0.19	    & 12.02$\pm$0.57\\
Swin-UNet       & 88.80$\pm$0.52	    & 12.75$\pm$0.95\\
MCTrans        & 89.50$\pm$0.32	    & 11.45$\pm$0.36 \\
{ BAT~\citep{BAT}}        & 89.87$\pm$0.28	    & \textbf{10.62$\pm$0.30} \\
UCTransNet       & 89.18$\pm$0.40	    & 12.37$\pm$0.61\\
\textbf{UDTransNet}   & \textbf{89.91$\pm$0.07}	    & {10.85$\pm$0.26}\\
\hline
\end{tabular}

\label{SOTA_COVID}
\end{table}

\begin{table}[]
\centering
\footnotesize
\caption{Quantitative results on the ACDC dataset in terms of average dice value (\%).}
\resizebox*{\linewidth}{!}{
\begin{tabular}{@{ }l|c|c@{\ \ }c@{\ \ }c@{ }}
\hline
Model        & Avg Dice & RV    & Myo   & LV    \\ \hline
R50-U-Net \citep{UNet}    & 87.55   & 87.10  & 80.63 & 94.92 \\
R50-AttnUNet \citep{AttentionUNet} & 86.75   & 87.58 & 79.20  & 93.47 \\
ViT-CUP \citep{ViT}      & 81.45   & 81.46 & 70.71 & 92.18 \\
R50-ViT-CUP \citep{ViT}  & 87.57   & 86.07 & 81.88 & 94.75 \\
TransUNet \citep{TransUNet}    & 89.71   & 88.86 & 84.53 & 95.73 \\
Swin-UNet \citep{SwinUnet}    & 90.00   & 88.55 & 85.62 & \textbf{95.83} \\
UNETR \citep{UNETR}        & 88.61   & 85.29 & \textbf{86.52} & 94.02 \\
UCTransNet \citep{UCTransNet}     & 89.69   & 87.92 & 85.43 & 95.71 \\ \hline
\textbf{UDTransNet}   & \textbf{90.11}       & \textbf{89.06} & 86.10 & 95.18     \\
\hline
\end{tabular}}
\label{SOTA_ACDC}
\end{table}

\textbf{ISIC-2018 Dataset for Lesion Segmentation.}
For skin lesion segmentation, ISIC-2018 dataset \citep{ISIC} was used in MICCAI'2018 Skin Lesion Analysis Towards Melanoma Detection Challenge. 
ISIC-2018 has a training set with 2594 images and their ground truth.
The 5-fold cross validation results of this dataset are reported, since the annotations of testing images are not publicly available.

\textbf{ACDC Dataset for Multi-class MRI Cardiac Segmentation.}
The Automated Cardiac Diagnosis Challenge (ACDC)~\citep{ACDC} was held at MICCAI'2017 which collects exams from different patients acquired from MRI scanners. Cine MRIs were acquired in a series of short-axis slices cover the heart from the base to the apex of the left ventricle. 
Each patient scan is manually annotated with ground truth for left ventricle (LV), right ventricle (RV) and myocardium (MYO). We report the average Dice of 5-fold cross validation results of the total 100 cases (1930 axial slices).

{
\textbf{COVID-19 Dataset for CT Lung Lesion Segmentation.}
COVID-19 CT scan lesion segmentation dataset\footnote{https://www.kaggle.com/datasets/maedemaftouni/covid19-ct-scan-lesion-segmentation-dataset} is a large dataset for COVID-19 by curating data from 7 public datasets. This dataset merges the COVID-19 lesion masks and their corresponding frames of these public datasets, with 2729 image and mask pairs. 
The 5-fold cross validation results are reported.}

\subsection{Implementation and Evaluation Details}
We trained our model with a single NVIDIA A40 GPU. Two kinds of online data augmentations are employed to avoid over-fitting, including random flipping and random rotating. 
We trained the models using Adam optimizer and the iteration number is not fixed due to the early-stopping strategy.
We use Cosine Annealing learning rate schedule \citep{Cosine}. 
We also choose the combined cross entropy loss and dice loss as our loss function for training, whereas dice coefficient (Dice) and 95\% Hausdorff distance (HD95) as the evaluation metrics for validating and testing. 
To make the results more stable and convincing, we conduct 5-fold cross validation, and obtain the mean value and standard deviation (std). 
The details of the implementation and some important hyper-parameters of the model are shown in Table~\ref{settings}.
Note that we use the same settings to train all the comparable models.

\begin{figure*}[!ht]
	\centering
	\includegraphics[width=\textwidth]{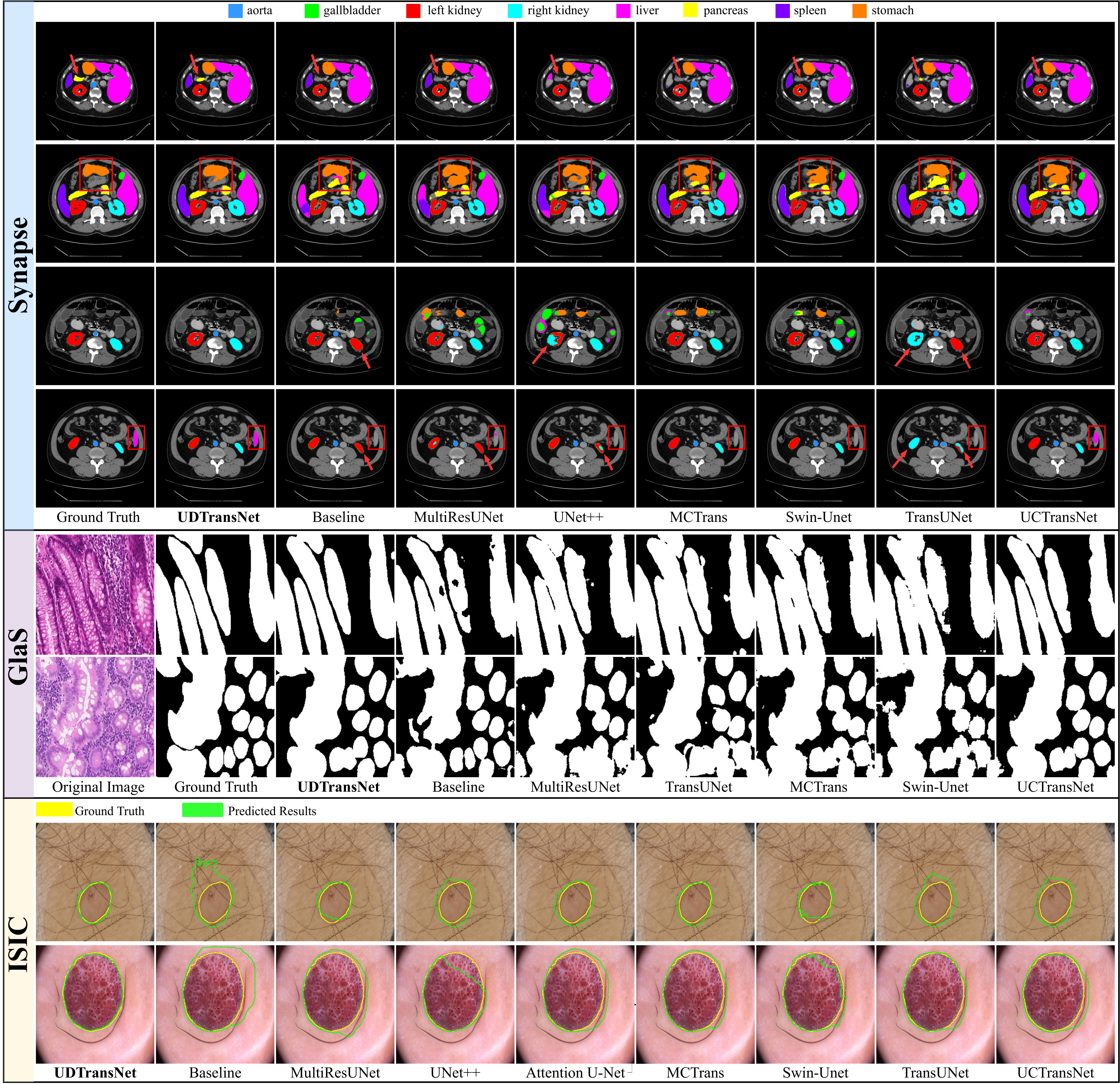} 
	\caption{The qualitative comparison of the segmentation results on three datasets by the state-of-the-art methods and the proposed UDTransNet.}
	\label{vis_SOTA_Synapse}
\end{figure*}

\subsection{Comparison to State-of-the-art Methods}

To demonstrate the strength of UDTransNet, we comprehensively compare UDTransNet with two categories of state-of-the-art segmentation methods, involving three CNN-based methods (UNet++, Attention U-Net and MultiResUNet) and five transformer-based methods (MedT, TransUNet, Swin-UNet, MCTrans and UCTransNet, which is the previous version of our model \citep{UCTransNet}). 
We also compared our methods with some task-specific state-of-the-art methods, e.g., BAT~\citep{BAT} which is carefully designed for the skin lesion segmentation tasks by utilizing transformers to capture the boundary information and solve the ambiguous boundaries of skin lesions, Inf-Net~\citep{InfNet} which is designed for the 
To make a fair comparison, their original released codes and settings are directly used in this experiment.

\subsubsection{Quantitative Results}

Quantitative results on the five datasets are reported in Table~\ref{SOTA_Synapse}, \ref{SOTA_GlaS}, \ref{SOTA_COVID} \& \ref{SOTA_ACDC} where the best results are boldfaced. Note that our results on the Synapse dataset is quite different with the results in \citep{TransUNet} and \citep{SwinUnet}, the reason is that their codes of evaluation metrics are not correct and we modified the code according to the rules of BraTS online evaluation\footnote{http://braintumorsegmentation.org/}.
The results demonstrate that our method obtains consistent improvements over the state-of-the-art methods.
Notably, our method also obtain better results than some task-specific state-of-the-art methods, e.g., BAT~\citep{BAT} in ISIC-18, Inf-Net~\citep{InfNet} in COVID-19, etc, which indicates that the proposed UDTransNet has more powerful generalization ability.
Our method also obtains relatively lower standard deviation which indicates that UDTransNet is more stable.
Compared with our previous algorithm UCTransNet, the current proposed UDTransNet also exhibits stronger capacity with more sophisticated attention modules.
Due to the intrinsic locality of convolution operations, CNN-based methods generally demonstrate limitations in explicitly modeling long-range dependency, resulting in inferior segmentation performance.
Compared with CNN-based methods, the transformer-based methods generally have better results on Synapse. 
Moreover, we found that the naive combination of CNN and transformer cannot produce satisfactory results, e.g., TransUNet (88.40\%) and MCTrans (88.44\%) gains inferior performance  than Attention U-Net (88.80\%) on GlaS whereas Swin-UNet (88.80\%) and UCTransNet (89.18\%) gain slightly lower dice scores than MultiResUNet (89.20\%) on ISIC. 
What is more, MedT always performs worse on all the datasets. However, our UDTransNet performs better than others with a relatively lower parameter number.
Our UDTransNet is boosted for two reasons: i) the fusion of the multi-scale features from the encoder stages; ii) the propagation of the appropriate features from the encoder to the decoder stage for balancing the possible semantic gaps. 
Moreover, the proposed collaborate learning framework and adaptive parameter selection scheme also contribute to the final performance.


\subsubsection{Qualitative Results}
Qualitative results of the comparable models are in Fig.~\ref{vis_SOTA_Synapse}.
The red boxes and arrows highlight the regions where UDTransNet performs better than the others.
The first four rows show the qualitative results on Synapse.
The results indicate that our model is better at capturing multi-scale objects. Besides, we observe that the small object of `pancreas' highlighted with red arrow (1st row) is missed by all the other methods. 
However, UDTransNet is able to segment it well thanks to the DAT module which enables feature aggregation of different resolutions.
It demonstrates that transmitting adequate spatial and semantic information from the encoder to the decoder helps identify the finer details by fusing the low-level spatial information.
Moreover, MCTrans, Swin-UNet and UDTransNet are able to distinguish the left kidney {from} the right kidney (3rd \& 4th rows), since the decoders obtain the global context through skip connections from the encoders. However, TransUNet fails to discriminate the similar regions since only the local features learned by convolution layers are transmitted to the decoder stage. This case also verifies that improving skip connections with a channel-spatial attention scheme is a more proper way, which facilitates the learning of the ambiguous objects.
As can be seen in the GlaS task (5th \& 6th rows), the segmentation results of UDTransNet are more similar to the ground truths than the other models. It not only eliminates the confusing false positive lesions but also produces coherent boundaries. 
These observations demonstrate that by fusing the low-level high-resolution and high-level semantic information, UDTransNet is capable of preserving detailed shape information for the segmentation task.
For the skin lesion segmentation of ISIC'18, the challenges are the shape variance and the boundary ambiguity. The under-segmentation and over-segmentation problems occur for the competing methods in the cases where the boundary is quite clear (7th row) or quite ambiguous (8th row). 
However, our UDTransNet is capable of achieving better performance in both cases thanks to the description of {segmented objects from a multi-scale perspective}.


\begin{table}[t]
\centering
\footnotesize
\caption{Ablation study of each component on GlaS and Synapse datasets in terms of Dice (\%). }
\begin{tabular}{@{}cccc|cc@{}}
\toprule
\multicolumn{4}{c|}{Components} & \multirow{2}{*}{GlaS(\%)} & \multirow{2}{*}{Synapse(\%)} \\
Baseline        & CFA           & SSA   & DRA   &                       &       \\\midrule
\checkmark  &     &       &      & \multicolumn{1}{c}{87.15$\pm$1.08}  & \multicolumn{1}{c}{86.05$\pm$0.64} \\

\checkmark  & \checkmark    &       &       & \multicolumn{1}{c}{89.62$\pm$0.77}  & \multicolumn{1}{c}{88.82$\pm$1.11} \\

\checkmark  &     &\checkmark       &       & \multicolumn{1}{c}{89.89$\pm$1.21}  & \multicolumn{1}{c}{88.90$\pm$0.46} \\

\checkmark  &\checkmark     &\checkmark   &    &        \multicolumn{1}{c}{90.28$\pm$0.72}  & \multicolumn{1}{c}{89.15$\pm$0.49} \\

\checkmark  & \checkmark    &\checkmark       &\checkmark       & \multicolumn{1}{c}{\textbf{91.03$\pm$0.56}}  & \multicolumn{1}{c}{\textbf{89.28$\pm$0.58}} \\ 

\bottomrule
\end{tabular}
\label{Ablation_GlaS}
\end{table}

\begin{figure}[t]
	\centering
	\includegraphics[width=0.8\linewidth]{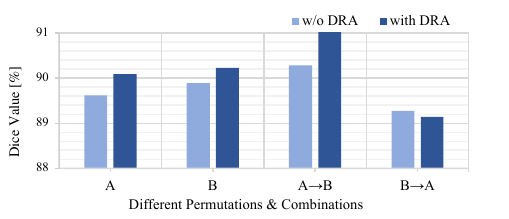} 
	\caption{Ablation study of different settings of the proposed DAT module {in} GlaS datasets. `A' denotes only CFA, `B' denotes only SSA and `A $\rightarrow$ B' is the proposed model.}
	\label{comb_permu}
\end{figure}

\subsubsection{Computational Complexity}
The computational complexity is an inherent disadvantage for transformers. However, our model is more parameter-efficient compared with the state-of-the-art transformer-based segmentation methods as shown in Table~\ref{SOTA_Synapse}.
Remarkably, the parameter amount of the UDTransNet is only 1/3 as the state-of-the-art transformer-based methods, \textit{e.g.} MedT, TransUNet and Swin-UNet, while achieving significant improvement.
Noting that TransUNet integrates the MSA module into the encoder-decoder framework whereas our UDTransNet incorporates the well-designed MSA module into the connections. It suggests the importance of design and the position of the MSA module in the U-shape architecture.

Moreover, the parameter number of UDTransNet is even lower than some CNN-based methods (UNet++, MultiResUNet, etc.). 
Furthermore, UDTransNet reduces the parameters of our previous model, UCTransNet, by half but improves the segmentation performance a lot. 
Except that the vanilla U-Net and U-Net with ResNet34 have fewer parameters but provide limited performance, our UCTransNet models achieve better segmentation performance with much fewer parameters as well as FLOPS and less inference time than other competing methods.
Considering the promising segmentation performance and the fewer parameters, it can be concluded that our model achieves a good trade-off between effectiveness and efficiency.

\subsection{Ablation Studies}
We evaluate the importance of each proposed component in UDTransNet through ablation experiments.
In the remainder of the paper, the U-Net with ResNet-34 as the backbone is chosen as the baseline model. 

\subsubsection{Ablation on the Proposed Components} 
As shown in Table~\ref{Ablation_GlaS}, compared to the baseline which achieves 87.71\% on the GlaS dataset, the dice scores increase to 89.62\% and 89.89\% when adding CFA and SSA, {respectively. The results} verify the effectiveness of multi-scale feature fusion and capturing inter- and intra-scale dependencies. 
The {consistent feature learned fused by} CFA provides more sufficient information for SSA than the simple concatenation of features from four scales, which increases the Dice to 90.28\% in the 4th row. This also proves that CFA and SSA are complementary.
Then, the DRA module is introduced to further increase the dice score to 91.03\%, which indicates the effectiveness of the decoder-guided feature recalibration.

\subsubsection{Ablation on {Orders} of the Proposed Components}
In this section, 
we systematically investigate different settings of the proposed DAT module to verify the {roles} of the three proposed attention-based modules in our UDTransNet.
As shown in Fig.~\ref{comb_permu}, when we change the order of CFA and SSA (`B$\rightarrow$A'), the dice scores decrease sharply, even worse than the result  with only  CFA or SSA. 
This observation demonstrates that it is necessary to fuse and align the channel information from the multi-scale stages 
with CFA before modeling the spatial relationship by  SSA for selecting the important regions. 
Furthermore, DRA is capable of further improving {performances} in most cases.

\begin{table}[t]
\centering
\footnotesize
\caption{Comparison between channel-wise attention (DRA-C) and spatial-wise attention (DRA-S) in DRA on the three datasets in terms of Dice (\%).}
\begin{tabular}{@{ }l|ccc@{ }}
\hline
Patterns          & GlaS      & Synapse    & ISIC   \\ 
\hline
DRA-C    & \textbf{91.03$\pm$0.56}    & \textbf{89.28$\pm$0.58}   & \textbf{89.91$\pm$0.07} \\
DRA-S    & 89.74$\pm$1.07    & 89.01$\pm$0.72   & 89.77$\pm$0.43 \\
\hline
\end{tabular}
\label{CS}
\end{table}

\begin{figure}[t]
	\centering
	\includegraphics[width=\linewidth]{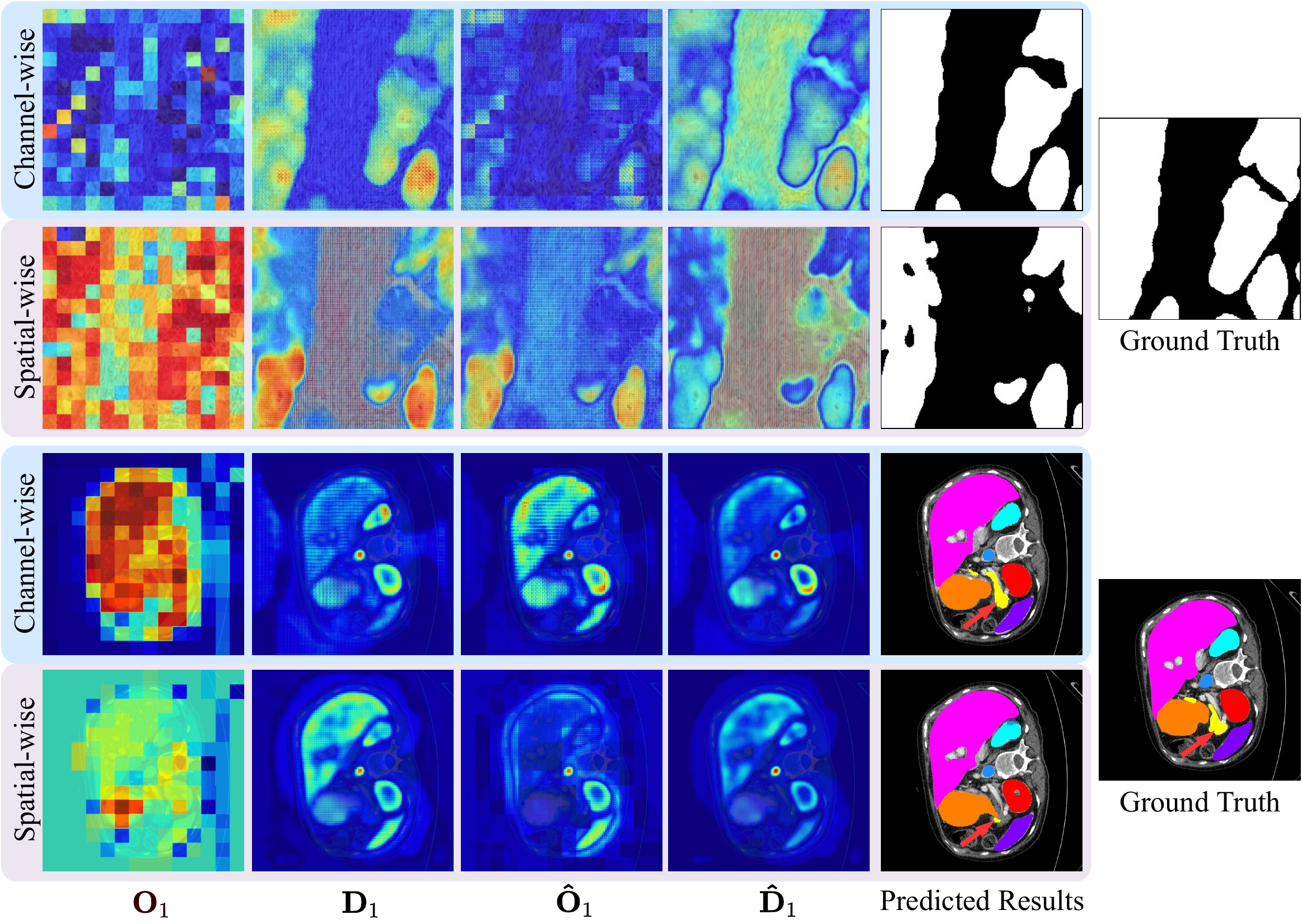} 
	\caption{Comparison of the feature maps between the DRA with channel-wise and spatial-wise attentions. }
	\label{vis_att}
\end{figure}

\subsubsection{Ablation on Different Attention Patterns of DRA}

We also test the performance of the DRA module with different attention mechanisms. Concretely, we compare our proposed channel-wise attention based DRA (DRA-C) with spatial-wise attention based DRA  (DRA-S).
As can be seen in Table~\ref{CS}, DRA-S performs worse than DRA-C in all the datasets, even worse than the results without DRA (90.28\% and 89.31\% on GlaS and Synapse), which indicates that {spatial attention leads to negative effects for fusing the incompatible features}.
To {investigate} the reason, we visualize the feature maps in the cases where DRA-S behaves poor. 
From Fig.~\ref{vis_att}, we observe that some useful patches of $\mathbf{O}_1$ are filtered out by DRA-S (e.g., 2nd row), leading to under-segmentation.
The channel-wise attention captures the more meaningful features to represent the target information whereas the spatial-wise attention highlights patches which are more important for all the feature maps.
Hence, the spatial attention {misleads the patch weighting and suppresses the critical patches when directly fusing the two semantically incompatible feature sets.}
Therefore, it is an appropriate way to bridge the semantic gap from the channel-wise perspective.

\section{Discussion}

\subsection{Computational Complexity}

The computational complexity is an inherent disadvantage for transformers. However, our model is more parameter-efficient compared with the state-of-the-art transformer-based segmentation methods as shown in Table~\ref{SOTA_Synapse}. For example, our model only has 33.8M parameters, but it consistently outperforms other models such as MedT, TransUNet and Swin-UNet which have more than 80M parameters. Furthermore, the parameter number of UDTransNet is even lower than some CNN-based methods (UNet++, MultiResUNet, etc.). 
Compared with our previous work UCTransNet, we reduce the width of the decoder and the embedding channels to achieve nearly half of the parameters.
Considering the promising segmentation performance and the fewer parameters, it can be concluded that our model achieves a good trade-off between effectiveness and efficiency.

\subsection{Influence of the Pre-trained Weights}

In the previous experiments, some of the transformer-based methods have pre-trained weights on the ImageNet, including TransUNet, Swin-UNet, MCTrans and our UDTransNet.
In this section, we trained those models from scratch and conduct a comparison to exploit behaviours of these methods without the pre-trained parameters.
As shown in Table~\ref{pretrain}, when trained from scratch, our UDTransNet has {increases} of 6.04\% (4.01\%), 9.60\% (4.63\%) and 5.18\% (5.98\%) on GlaS (Synapse) dataset compared with TransUNet, Swin-UNet and MCTrans in terms of Dice, respectively. 
These existing transformer-based models suffer 3.3\%, 8.2\%, and 3.7\% drop on Dice when trained from scratch, even worse than CNN-based segmentation models.
The results indicate that TransUNet and Swin-UNet highly rely on the learned initial features 
whereas UDTransNet has more powerful learning ability with fewer parameters, which further suggests that redesigning the skip connection with transformer is a more proper way of taking full advantage of the U-Net and transformer.

\subsection{Influence of the Backbones}
In addition, we also verify the influence of the backbones on the segmentation performance. 
It is worth noting that the backbones of the several competing methods are specifically tailored, such as the pure transformer architecture in Swin-UNet for Swin Transformer. When the backbone is replaced in one of the comparable methods, e.g., TransUNet or Swin Transformer, the performance drops significantly.
As shown in Table~\ref{backbone}, when replaced with the same backbones, our UDTransNet still {achieves} better performances except with Swin Transformer as backbone, which indicates that the fine-grained convolutional features are necessary when combined with the long-range modeling based the MSA modules. 
How to integrate the two operations, i.e. convolution and MSA, is crucial for enhancing the segmentation performance.
It again indicates that integrating the MSA components into the skip-connection is more appropriate than the encoder or decoder stage.

\subsection{Sensitivity to the Numbers of Heads and Layers}
\begin{figure}[t]
	\centering
	\includegraphics[width=\linewidth]{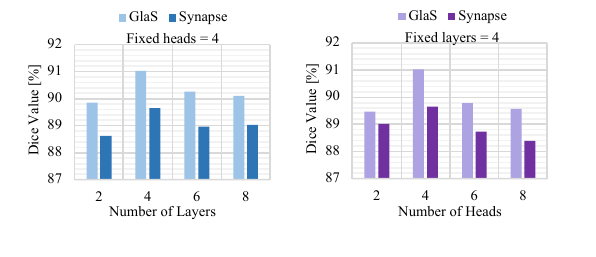} 
	\caption{Sensitivity to the Numbers of Layers and Heads of DAT Module.}
	\label{Num_L_H}
\end{figure}

In this section, we investigate the sensitivity of our model with respect to its two hyper-parameters: the number of heads $N_H$ and the number of layers $N_L$ in the proposed DAT module.
We first keep the number of heads $N_H$ of SSA fixed and gradually increase the number of layers $N_L$ of DAT to explore the optimal $N_L$. Then we keep $N_L$ fixed and change $N_H$.
As shown in Fig.~\ref{Num_L_H}, the optimal numbers of heads and layers are $N_H = 4$ and $N_L = 4$, which demonstrates that more heads and layers help model multi-scale representations but too many heads and layers introduce redundant parameters to the model, leading to over-fitting as well.
This implies that {improving} the capacity of transformer-based model is not appropriate regardless of small datasets or large datasets.
Moreover, the variations of the different numbers of layers and heads are not large, which indicates that our UDTransNet is not sensitive to the numbers of layers and heads.

\begin{table}[t]
\centering
\footnotesize
\caption{Experiment results of different transformer-based methods without pre-trained weights in terms of Dice (\%).}
\resizebox*{\linewidth}{!}{
\begin{tabular}{@{ }lcccc@{ }}
\hline
Method          & Backbone      & Pre-train & GlaS     & Synapse \\ 
\hline
\multirow{2}{*}{TransUNet}       & \multirow{2}{*}{ResNet-50}     & R50+ViT   &87.41$\pm$1.15	 &87.21$\pm$0.88\\ & & From scratch   &84.49$\pm$1.04	 &85.30$\pm$1.48\\ \hline 
\multirow{2}{*}{Swin-UNet}       & \multirow{2}{*}{Swin}          & Swin   &89.04$\pm$0.88	 &86.73$\pm$0.67\\ & & From scratch  &81.75$\pm$2.52 	 &84.79$\pm$1.43\\ \hline
\multirow{2}{*}{MCTrans}        & \multirow{2}{*}{ResNet-34}   & R34      & 88.44$\pm$0.74	    & 87.65$\pm$0.81\\ & & From scratch   & 85.18$\pm$1.55	    & 83.71$\pm$2.26\\ \hline
\multirow{2}{*}{\textbf{UDTransNet}}   & \multirow{2}{*}{ResNet-34}     & R34 &\textbf{91.03$\pm$0.56}	 &\textbf{89.28$\pm$0.58}\\ & & From scratch &\textbf{89.59$\pm$1.15}	 &\textbf{88.72$\pm$0.58}\\
\hline
\end{tabular}}
\label{pretrain}
\end{table}

\begin{table}[t]
\centering
\footnotesize
\caption{Comparison of our UDTransNet with the same backbones as the transformer-based methods in terms of Dice (\%).}
\begin{tabular}{@{ }lccc@{ }}
\hline
Method          & Backbone    & GlaS     & Synapse \\ 
\hline
{TransUNet}       & {ResNet-50}     &87.41$\pm$1.15	 &87.21$\pm$0.88\\ 
{{UDTransNet}}   & {ResNet-50}    &\textbf{90.45$\pm$0.61}	 &\textbf{89.97$\pm$0.45}\\ \hline 
{Swin-UNet}       & {Swin Transformer}          &\textbf{89.04$\pm$0.88}	 &86.73$\pm$0.67\\  
{{UDTransNet}}   & Swin Transformer   &{88.03$\pm$1.27}	 &\textbf{87.66$\pm$0.68}\\ \hline
{MCTrans}        & {ResNet-34}      & 88.44$\pm$0.74	    & 87.65$\pm$0.81\\
{{UDTransNet}}   & {ResNet-34}    &\textbf{91.03$\pm$0.56}	 &\textbf{89.28$\pm$0.58}\\ 
\hline
\end{tabular}
\label{backbone}
\end{table}

\section{Conclusion}

Accurate and automatic segmentation of medical images is a crucial step for clinical diagnosis and analysis.
However, there {are relatively fewer studies} on incorporating the attention mechanism into the skip connection. 
In this work, we proposed a transformer-based segmentation framework (UDTransNet) by exploring the channel- and spatial-wise attention mechanisms on the connections between encoder-decoder stages for precise medical image segmentation.
By combining the strengths of multi-scale Dual Attention Transformer (DAT) and the Decoder-guided Recalibration Attention (DRA), the proposed approach achieves state-of-the-art results on multiple  medical image segmentation tasks, covering abdominal organs segmentation, skin lesion segmentation and gland nuclei segmentation.
Through {thorough comparisons} and in-depth analysis, we verified the effectiveness of each component in our UDTransNet, which successfully narrows the semantic gaps and fuses the multi-scale features with different attention mechanisms.
Our results shed new light on the importance of the attention mechanism for enhancing the skip connections in the segmentation tasks. 
We also obtain some interesting findings regarding the interpretability of channel- and spatial-wise attentions which  are important research topics for {the} future studies.

\bibliographystyle{model2-names.bst}\biboptions{authoryear}
\bibliography{refs}

\end{document}